\documentclass[journal,twoside,web]{IEEEtran}
\usepackage{cite}
\usepackage{amsmath,amssymb,amsfonts}
\usepackage{multirow}
\usepackage{algorithmic}
\usepackage{graphicx}
\usepackage{textcomp}
\usepackage{hyperref}
\usepackage{subfig}
\def\BibTeX{{\rm B\kern-.05em{\sc i\kern-.025em b}\kern-.08em
		T\kern-.1667em\lower.7ex\hbox{E}\kern-.125emX}}
\begin{document}
	\title{Supervised Segmentation with Domain Adaptation for Small Sampled Orbital CT Images}
	\author{Sungho Suh, Sojeong Cheon, Wonseo Choi, Yeon Woong Chung, Won-Kyung Cho, Ji-Sun Paik,\\ Sung Eun Kim, Dong-Jin Chang, and Yong Oh Lee 
		\thanks{For CMC-ORBIT dataset, the central Institutional Review Board (IRB) of Catholic Medical Center approval was obtained (XC19REGI0076). S. Suh and S. Cheon contributed equally to this work. (Corresponding authors: Y. O. Lee, email:yongoh.lee@kist-europe.de) }
	
	}
	
	\maketitle
	
	\begin{abstract}
		Deep neural networks (DNNs) have been widely used for medical image analysis. However, the lack of access a to large-scale annotated dataset poses a great challenge, especially in the case of rare diseases, or new domains for the research society.  Transfer of pre-trained features, from the relatively large dataset is a considerable solution. In this paper, we have explored supervised segmentation using domain adaptation for optic nerve and orbital tumor, when only small sampled CT images are given. Even the lung image database consortium image collection (LIDC-IDRI) is a cross-domain to orbital CT, but the proposed domain adaptation method improved the performance of attention U-Net for the segmentation in public optic nerve dataset and our clinical orbital tumor dataset. The code and dataset are available at \url{https://github.com/cmcbigdata}.
	\end{abstract}
	
	
	\section{Introduction}
	\label{sec:introduction}
	
	\IEEEPARstart{D}{eep} Neural networks have made great breakthroughs in various computer vision tasks such as image classification, semantic and instance segmentation, object detection, and so on. One of the main contributions to these achievements, is the large-scale dataset with detailed annotations. For instance, ImageNet dataset \cite{deng2009imagenet} consisting of more than 14 million images, equipped with more than 20 thousand classes, is a golden standard open dataset for image classification, and MS COCO dataset \cite{lin2014microsoft} collecting more than a million images with instance segmentation annotations for object segmentation. Unlike the natural image domain, it is challenging to collect a sufficiently large amount of dataset in the medical image domain. Annotated medical datasets are limited due to the laborious labeling process by the trained experts, and legal issues associated with publicly sharing private medical information.
	
	Many efforts have been made to obtain open-access medical image datasets. 
	However, the amount of data is much smaller as compared to other open datasets, also targeted diseases are biased towards well-known artificial intelligence related research societies. For instance, the open dataset of retinal fundus images can be found in the ophthalmology domain because a deep learning-based diagnosis of diabetic retinopathy is a well-known topic \cite{ting2019artificial}. In the case of the orbital tumor, open access data is rarely found in the ophthalmology dataset. Orbit is the space that surrounds an eyeball and soft tissue, and contains many elements that are vital for vision. Tumors arising within or around the orbit would compress vital structures for vision, and may directly threaten the vision because the orbit is composed of rigid bones. The differentiation of the orbital tumor from benign and malignant, through deep learning analysis, will innovatively change the field of medical treatment, especially for the challenging cases of biopsy trail. Other rare diseases, apart from ophthalmology have also struggled with the same issue.
	
	Data augmentation and transfer learning/domain adaptation have been proposed to tackle the scarcity of medical images for training deep neural network (DNN) models. The data augmentation approach involves training a model using synthetic data generated by generative adversarial networks (GANs). 
	
	The GAN frameworks have been employed to create label-to-segmentation translation, segmentation-to-image translation or medical cross-modality translation \cite{yi2017dualgan, liu2017unsupervised, choi2018stargan}. The method for synthetic data augmentation, used to enlarge medical dataset was able to improve medical image detection \cite{han2019combining}, classification \cite{frid2018gan}, and segmentation \cite{han2019synthesizing}. However, the improvement was limited due to transferring images, which can be biased by the small-scale training dataset, or distorted at the pixel level in the localized volume of interests \cite{cao2020auto}. 
	
	The transfer learning/domain adaptation approach involves capturing the features from the source domain where the amount of data is quite large, and fine-tuning the pre-trained features in the target domain.

	This feature adaptation process provides more accurate predictions for the target domain. Regardless of the apparent difference between source and target domains, most of these methods discriminate feature distributions of source and target domains in adversarial learning \cite{chen2019synergistic}. Although, the adversarial discriminators implicitly enhance features extracted using DNNs, the adaptation process can outperform in the segmentation of medical images, where only a small number of samples are available.
	
	In this paper, we have proposed a supervised segmentation with domain adaptation for small sampled orbital CT images. The supervised segmentation model is a 2D sequential U-Net with attention combined using Sensor3D \cite{novikov2018deep} and Attention U-net \cite{oktay2018attention}. Domain adaptation methods were employed for the segmentation of optic nerve and orbital tumor from about 50 and 20 samples of orbital CT scans. 
	
	Additionally, the adversarial learning between the segmentation model and discriminator was applied while domain adaptation was performed.
	
	The main contributions of this paper are (1) we have proposed simple domain adaptation methods that require only a small amount of annotated data in the target domain, (2) we performed domain adaptation from lung nodule CT images to orbital CT images that were in the same modality, but in different disease domains, (3) we performed an ablation study for evaluating the effect of domain adaptation, and (4) to our best knowledge, this is the first attempt at orbital tumor segmentation based on deep learning, and providing an orbital tumor dataset to the public. 
	
	The paper is organized as follows. In Section \ref{sec:relatedwork}, we have presented related work. Section \ref{sec:proposedmethod} describes the base 2D sequential U-Net in detail, and the proposed domain adaptation methods. In section \ref{sec:experimentalresults}, we described the dataset used in our experiments and the experiment results, followed by an ablation study. Finally, in section \ref{sec:conclusions}, we discussed the findings and limitations of our study. Supplements are provided for more detailed data descriptions of orbital tumor dataset and source code that are available at \url{https://github.com/cmcbigdata}.

	\section{Related Work}
	\label{sec:relatedwork}
	
	The goal of medical image segmentation is to classify the pixels in an image, thereby identifying internal organs, and recognizing abnormal areas like tumors and lesions. Convolutional neural network (CNN) based approaches, such as fully convolution networks \cite{long2015fully}, Deeplab \cite{chen2017deeplab}, and U-Net \cite{ronneberger2015u}, have researched and achieved great success in medical image segmentation tasks. The FCN predicts a dense output matrix size as used in the original input, and has an encoder-decoder structure. Due to the success of the FCN in natural imaging \cite{long2015fully}, various types of the encoder-decoder networks with U-shaped architectures, such as SegNet \cite{badrinarayanan2017segnet}, U-Net \cite{ronneberger2015u}, and U-Net++ \cite{zhou2018unet++}, have been proposed. These network models have been proposed for medical image segmentation of different anatomical organs, such as liver and liver tumor \cite{li2018h, seo2019modified, xi2020cascade}, brain and brain tumor \cite{menze2014multimodal, dai2018automatic, akil2020fully}, lung and lung nodule \cite{jin2018ct, aresta2019iw}, polyp \cite{akbari2018polyp, zhang2019real}, etc.
	
	Based on CNN, the U-Net architecture composes of an encoder, a decoder, and skipped connections between the encoder and decoder. Upsampled feature maps are summed with feature maps skipped from the encoder in the FCN, while the U-Net concatenates them and adds convolutions and non-linearities between each upsampling step. The encoder includes downsampling to extract the context information, while the decoder is an upsampling process that combines the upsampled features and the low-dimensionality features from the downsampling layer, to improve the network performance. However, the U-Net architecture has some limitations, such as the lack of flexibility in the structure of the model when it is trained with differently sized datasets. Further, the skip connection does not fully exploit the features from the encoder. To improve the structure of the U-Net for medical image segmentation, recent studies have proposed redesigning the skip connection structure, changing the structure of the encoder and decoder, or using a dense cascade structure.
	
	Zhou et al. \cite{zhou2018unet++} proposed an UNet++ model, which exploited the multiscale features by redesigning the nested skip connections. The redesigned skip connections included the convolution units that were connected as a dense network. Oktay et al. \cite{oktay2018attention} proposed an attention U-Net model for medical imaging, which could learn to concentrate on target structures of different shapes and sizes. Chen et al. \cite{chen2018drinet} proposed Dense-Res-Inception Net (DRINet) for medical image segmentation. The DRINet combined a convolution block using dense connection \cite{huang2017densely} and a deconvolutional block using residual inception modules \cite{szegedy2015going, he2016deep}. Although, these methods improved the performance of the U-Net-based network for medical image segmentation, 3D convolutional medical image segmentation still requires high computational costs. 
	
	Inspired by the DenseNet architecture \cite{huang2017densely}, Li et al. \cite{li2018h} proposed H-DenseUNet, which combined a slice-wise densely connected variant of the 2D and 3D U-Net architecture. The 2D U-Net architecture extracted the intra-slice features in each 2D image from a 3D CT scan, followed by which, the 3D U-Net architecture aggregated the volumetric contexts and refined the result using the auto-context algorithm in 3D. However, H-DenseUNet is complex and consumes a large memory. Additionally, it tends to fuse semantically dissimilar feature maps from the encoder and decoder sub-networks, which can degrade segmentation performance.
	
	Another hybrid approach that combines the U-Net structure with recurrent networks, such as convolutional long short term memory (C-LSTM) \cite{shi2015convolutional}. 
	Sensor3D \cite{novikov2018deep} is a deep sequential segmentation model that integrated a 2D U-Net and two bidirectional C-LSTM networks, one at the latent space and the other at the last decoder layer.
	The segmentation by the Sensor3D was performed by feeding only three adjacent slices as input. The limitation of large memory and whole volume requirement in the conventional 3D U-Net architecture can be released by learning the spatial and temporal relationships, using only the image and its two adjacent ones.
	
	In ophthalmology, various deep learning models have been applied to ocular imaging, principally fundus photographs and optical coherence tomography. Deep learning-based detection and segmentation models are used for  diabetic retinopathy \cite{o11, o12, o13}, glaucoma \cite{o11, o16}, age-related macular degeneration \cite{o11, o18} and retinopathy of prematurity \cite{o19}. These ophthalmic diseases have publicly available big data.
	
	\begin{figure*}
		\centerline{\includegraphics[width=2\columnwidth]{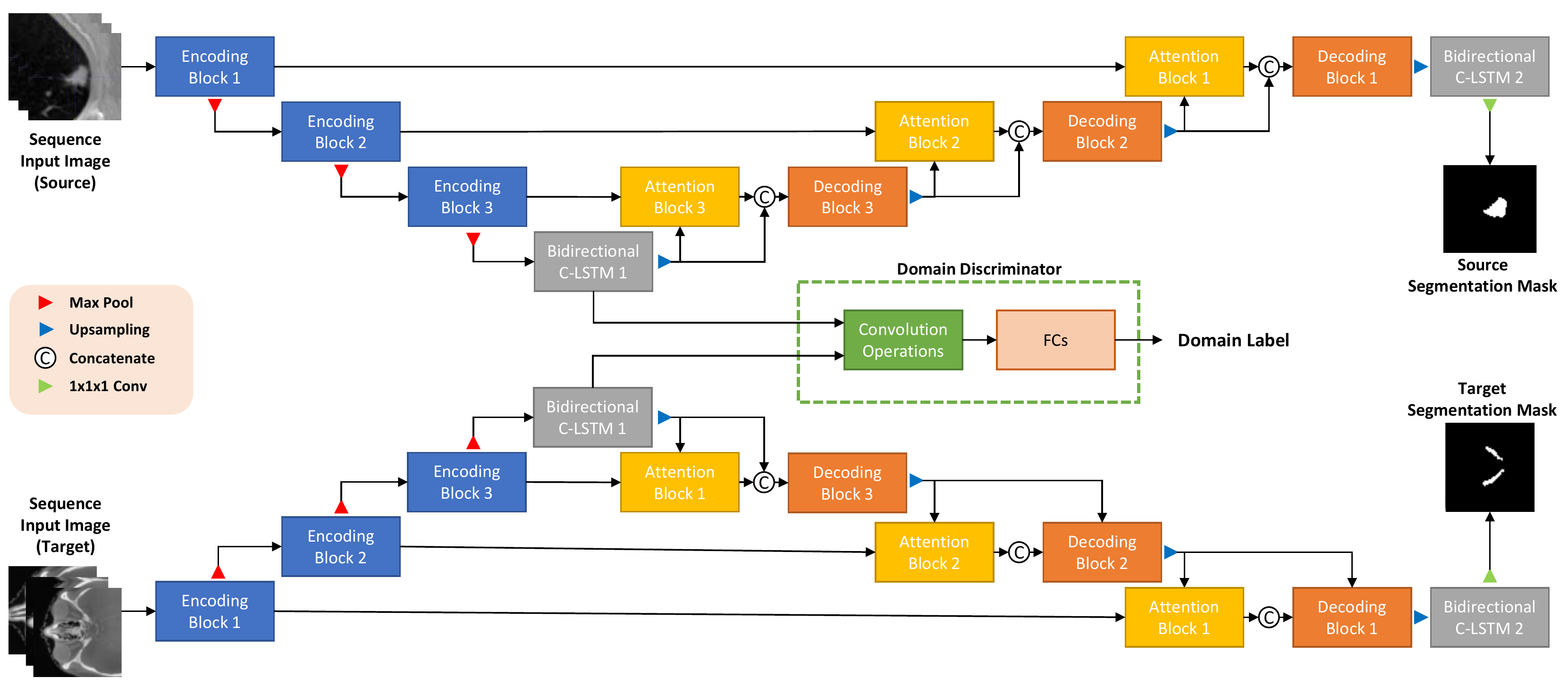}}
		\caption{The overall network architecture of the proposed framework with domain adaptation}
		\label{fig:overview}
	\end{figure*}
	
	Unfortunately, the conventional image segmentation models require large and high-quality annotated datasets, but the common limitations of such medical datasets are that both, data and annotations are expensive to acquire, while the annotations are also sparse and noisy. Recently, the problem of the small annotated datasets has been alleviated by employing a large labeled source dataset via transfer learning. The segmentation of orbital CT images is one of the ophthalmic diseases whose publicly available large and annotated datasets are a rare find.
	Transfer learning is a machine learning method that transfers knowledge learned in a source task to a target task \cite{pan2009survey}. The weights of DNNs are first pre-trained on a large-scale dataset, which is called the source task, and then fine-tuned using the data from the target task, using a small amount of data \cite{pan2009survey}. Domain adaptation, a subclass of transfer learning, attempts to bridge the distribution gap between source and target tasks by either learning a common latent representation, or by learning to translate images from the source domain to the target domain. GAN \cite{goodfellow2014generative} and its variants such as CycleGAN \cite{zhu2017unpaired} are generally integrated into the domain adaptation techniques in the same manner. 
	
	GAN and domain adaptation have been utilized in the medical image analysis. Zhang et al. \cite{da58} proposed synthesizing and segmenting multimodal medical volumes, using GANs with the cycle and shape consistencies. In \cite{da59}, a domain adaptation module that maps the target input to features that are aligned with source domain feature space, using a domain critic module for discriminating the feature space of both domains, was proposed for cross-modality biomedical image segmentation. Huang et al. \cite{da60} proposed a universal U-Net comprising of domain-general and domain-specific parameters to deal with multiple organ segmentation tasks on multiple domains. This integrated learning mechanism offers a new possibility of dealing with multiple domains, as well as multiple heterogeneous tasks. However, most studies are based on cross-modality image analysis. This paper considers a method of domain adaptation for cross-disease in a single domain, CT.

	\section{Proposed Method}
	\label{sec:proposedmethod}
	
	In this section, we introduce the details of the proposed framework for orbital CT scan image segmentation. The proposed model is composed of two components: an attention U-Net architecture with bidirectional C-LSTMs, and a U-Net architecture-based domain adaptation. The attention U-Net architecture with bidirectional C-LSTMs, denoted by \textit{SEQ-UNET}, was inspired by \cite{oktay2018attention} and \cite{novikov2018deep}. To reduce computational cost of 
	3D convolution operations, we considered a 3D volumetric CT scan as a sequence of 2D slice images. The sequential CT slices are input into the model.
	To conserve spatio-temporal correlation between 2D slices, two bidirectional C-LSTMs were employed at the end of the encoder and decoder.
	We employed attention gates between the encoder and decoder structures in the proposed U-Net model to highlight salient features for orbital CT scan image segmentation. 
	 
	Furthermore, we applied the domain adaptation approach to the feature extraction process, using an adversarial learning framework. Fig. \ref{fig:overview} displays the overall procedure of the proposed framework. 
	
	\subsection{Attention U-Net with bidirectional C-LSTMs: SEQ-UNET}
	
	A 3D volumetric CT scan was considered as a sequence of 2D slice images. A general approach of handling sequential data extract features from unidirectional temporal correlations between image frames, was opted. Unlike the general sequential data, the 2D slice images were not real sequential data and required learning the inter-slice correlations, from bidirectional spatio-temporal correlations. To preserve the order of slices and the correlation between the slices, we employed bidirectional C-LSTMs in a U-Net architecture, inspired from \cite{novikov2018deep}. The bidirectional C-LSTMs in the proposed model were based on the version of the LSTM without connections between the cell and the gates, to reduce computational costs. 
	
	U-Net \cite{ronneberger2015u} concatenate feature maps from different levels to improve image segmentation performance, and combines the low-level detail information with high-level semantic information. The U-Net architecture is composed of an encoder, a decoder, and skipped connections between the two. The encoder includes several convolution layers that extract increasingly abstract representations of the input image, and max-pool downsampling layers to encode the input image into feature representations, at multiple different levels. Similarly, the decoder is also equipped with several convolution layers, upsampling layers, and concatenation blocks. The decoder semantically projected the discriminative features, learned by the encoder onto a higher resolution image space to obtain a dense classification. The extracted features from the encoder in the proposed U-Net architecture were fed into the bidirectional C-LSTM to capture spatio-temporal correlations between 2D image slices of a 3D CT scan. The bidirectional C-LSTM equipped with the extracted features from the encoder extracted the order, and inter-slice correlations from the high-level semantic information of the slice images. Additionally, the output of the decoder was also input into the bidirectional C-LSTM. The bidirectional C-LSTM along with the output of the decoder added explicit dependency for the low-level detailed information, and converted the dimensions of the slice sequence into a one-dimensional image, same as a segmentation mask image.
	
	\begin{table*}
		\caption{Detailed structure of SEQ-UNET + DA 
		when the length of input sequence is 3 and the size of input slice image is 64 $\times$ 64. The fourth column indicates the activation function used in the layer. "$d \times$ []" represents the convolution layer that is repeated for $d$ times.}
		\label{tab:networkarchitecture}
		\centering
		\resizebox{2\columnwidth}{!}{
			\begin{tabular}{c|c|c|c|c|c}
				\hline
				Network & Name & Layers & Act. Func. & Input Tensor & Output Dimension\\
				\hline
				\multirow{8}{*}{Feature Extractor} 
				& Input Sequential Images & - & - & - & 3 $\times$ 1 $\times$ 64 $\times$ 64\\ 
				& Encoding Block 1	& 2 $\times$ [Conv 3 $\times$ 3, st=1] & ELU & Input & 3 $\times$ 64 $\times$ 64 $\times$ 64\\
				& Pooling 1			& Max Pooling 2 $\times$ 2, st=2 	 & - & Encoding 1	& 3 $\times$ 64 $\times$ 32 $\times$ 32\\ 
				& Encoding Block 2	& 2 $\times$ [Conv 3 $\times$ 3, st=1] & ELU & Pooling 1	& 3 $\times$ 128 $\times$ 32 $\times$ 32\\
				& Pooling 2			& Max Pooling 2 $\times$ 2, st=2 	 & -   & Encoding 2 & 3 $\times$ 128 $\times$ 16 $\times$ 16\\
				& Encoding Block 3	& 2 $\times$ [Conv 3 $\times$ 3, st=1] & ELU 	& Pooling 2 & 3 $\times$ 256 $\times$ 16 $\times$ 16\\
				& Pooling 3			& Max Pooling 2 $\times$ 2, st=2 	 & -	& Encoding 3 & 3 $\times$ 256 $\times$ 8 $\times$ 8\\ 
				& Bidirectional C-LSTM 1 & Bidirectional C-LSTM 3 $\times$ 3 & - & Pooling 3 & 3 $\times$ 512 $\times$ 8 $\times$ 8\\
				\hline			
				\multirow{4}{*}{Domain Discriminator}
				& Summation 1 	& Pixel summation & - & Bidirectional C-LSTM 1 & 512 $\times$ 8 $\times$ 8\\
				& Conv 1 		& 3 $\times$ [Conv 5 $\times$ 5, st=2] & Leaky ReLU & Summation 1 	& 128 $\times$ 1 $\times$ 1\\
				& FC 1 			& Fully Connected					   & ReLU		& Conv 1 		& 10 \\
				& FC 2 		  	& Fully Connected 					   & - 		    & FC 1			& 2 \\	
				\hline
				\multirow{17}{*}{\shortstack{Reconstructor with \\ Attention Gates}}
				& Upsampling 1 & Upsampling 2 $\times$ 2 & - & Bidirectional C-LSTM 1 & 3 $\times$ 512 $\times$ 16 $\times$ 16\\
				& \multirow{2}{*}{Attention Block 3} & \multirow{2}{*}{Conv 1 $\times$ 1, st=1, Eq.(\ref{eq:attention})} & \multirow{2}{*}{ReLU \& Sigmoid} & Upsampling 1, & 3 $\times$ 256 $\times$ 16 $\times$ 16\\
				&									 &  & & Encoding 3 & 3 $\times$ 256 $\times$ 16 $\times$ 16\\
				& Concatenate 1 & Concatenate & - & Upsampling 1, Attention 3 & 3 $\times$ 768 $\times$ 16 $\times$ 16\\
				& Decoding Block 3 & 2 $\times$ [Conv 3 $\times$ 3, st=1]	& ELU & Concatenate 1 & 3 $\times$ 256 $\times$ 16 $\times$ 16\\
				& Upsampling 2	& Upsampling 2 $\times$ 2 & - & Decoding Block 3 & 3 $\times$ 256 $\times$ 32 $\times$ 32\\ 
				& \multirow{2}{*}{Attention Block 2} & \multirow{2}{*}{Conv 1 $\times$ 1, st=1, Eq.(\ref{eq:attention})} & \multirow{2}{*}{ReLU \& Sigmoid} & Upsampling 2, & 3 $\times$ 128 $\times$ 32 $\times$ 32\\
				&									 &  & & Encoding 2 & 3 $\times$ 128 $\times$ 32 $\times$ 32\\
				& Concatenate 2 & Concatenate & - & Upsampling 2, Attention 2 & 3 $\times$ 384 $\times$ 32 $\times$ 32\\
				& Decoding Block 2 & 2 $\times$ [Conv 3 $\times$ 3, st=1]	& ELU & Concatenate 2 & 3 $\times$ 128 $\times$ 32 $\times$ 32\\
				& Upsampling 3	& Upsampling 2 $\times$ 2 & - & Decoding Block 2 & 3 $\times$ 128 $\times$ 64 $\times$ 64\\ 
				& \multirow{2}{*}{Attention Block 1} & \multirow{2}{*}{Conv 1 $\times$ 1, st=1, Eq.(\ref{eq:attention})} & \multirow{2}{*}{ReLU \& Sigmoid} & Upsampling 3, & 3 $\times$ 64 $\times$ 64 $\times$ 64\\
				&									 &  & & Encoding 1 & 3 $\times$ 64 $\times$ 64 $\times$ 64\\
				& Concatenate 3 & Concatenate & - & Upsampling 3, Attention 1 & 3 $\times$ 192 $\times$ 64 $\times$ 64\\
				& Decoding Block 1 & 1 $\times$ [Conv 3 $\times$ 3, st=1]	& ELU & Concatenate 3 & 3 $\times$ 64 $\times$ 64 $\times$ 64\\
				& Bidrectional C-LSTM 2 & Bidirectional C-LSTM 3 $\times$ 3 & - & Decoding 1 & 1 $\times$ 64 $\times$ 64 $\times$ 64\\
				& Segmentation Output	& Conv 1$\times$ 1, st=1 & Sigmoid & Bidirectional C-LSTM 2 & 1 $\times$ 1 $\times$ 64 $\times$ 64\\
				\hline		
		\end{tabular}}
	\end{table*}
	
	As mentioned prior, the U-Net architecture comprises of three blocks, where the bidirectional C-LSTMs were further applied to improve the performance of the encoder and decoder. In the proposed model, we installed attention gates before the skipped connections to highlight salient features, and merge only relevant activations. The attention gates filtered the neuron activations during the forward pass, as well as during the backward pass. As suggested in \cite{oktay2018attention}, we used additive attention \cite{bahdanau2014neural} to obtain the gating coefficient that contained contextual information, to prune lower-level feature responses. In the U-Net architecture, the gating vector implies the feature from the convolution layer in the decoder, and the input for the attention gate implies the feature extracted from the convolution layer in the encoder. Additive attention is expressed as follows:
	
	\begin{equation}
		\label{eq:attention}
		\begin{split}
			q^l_{att} = \psi^T (\text{ReLU}(W_x^T x_i^l + W_g^T g_i + b_g)) + b_{\psi}\\
			\alpha^l_i = \text{Sigmoid}(q^l_{att}(x^l_i , g_i; \Theta_{att}))
		\end{split}
	\end{equation}
	where, $x_i^l$ denotes the pixel vector in the feature map in layer $l$, $g$ denotes the gate vector, $W_x$ and $W_g$ denote linear transformations, and $b_g$ and $b_{\psi}$ are bias terms. By adding the attention gates before the skip connections, information extracted from the coarse scale was used in gating, to disambiguate irrelevant and noisy responses in skip connections. The detailed structure of the proposed attention U-Net equipped with bidirectional C-LSTMs is shown in Table \ref{tab:networkarchitecture}. 
	
	\subsection{Domain Adaptation with Domain Discriminator: SEQ-UNET + DA}
	
	To alleviate the problem of small annotated datasets, we applied the domain adaption technique to SEQ-UNET.
	The feature representation learning process extracted features that were more invariant to differences between the two domains, by employing a domain discriminator to distinguish between the source and target features from the encoder, in the U-Net architecture. Fig. \ref{fig:overview} presents the overall network architecture of the proposed framework with the domain adaptation.
	
	There are four types of network structures in the proposed framework: 1) a feature extractor inclusive of encoding blocks in the U-Net structure, and a bidirectional C-LSTM, 2) attention gates, 3) a reconstructor, including decoding blocks in the U-Net structure and a bidirectional C-LSTM, and 4) a domain discriminator. The domain discriminator was appended to the end of the encoder with the bidirectional C-LSTM layer, with the most complex feature. We used adversarial learning that had an objective function of the min-max game between the feature extractor and the domain discriminator. The feature extractor extracted domain invariant features to trick the domain discriminator, while the domain discriminator was trained to distinguish between features extracted from the source and target domain data. The detailed structure of the proposed domain adaptation networks is also shown in Table \ref{tab:networkarchitecture}. Furthermore, we integrated a dice loss function into the overall loss function, for the segmentation loss between the segmentation mask and the result from the reconstructor. This was done because the data ratio between the background and foreground was severely imbalanced, and the dice score could effectively mitigate the problem of class imbalance. The objective functions of the proposed method are defined as follows:
	
	\begin{equation}
		\label{eq:domainadaptation_discriminator}
		\begin{split}
			\mathop{\mathbb{L}_D}(x_{src},x_{tgt};\theta_D) = &\mathop{\mathbb{L}_{CE}}(D(E(x_{src})), 0) \\
			&+ \mathop{\mathbb{L}_{CE}}(D(E(x_{tgt})), 1)
		\end{split}
	\end{equation}
	
	\begin{equation}
		\label{eq:domainadaptation_segmentation}
		\begin{split}
			\mathop{\mathbb{L}_{Seg}}(x_{src},x_{tgt},m_{tgt};\theta_{Seg}) &= \mathop{\mathbb{L}_{CE}}(D(E(x_{tgt})), 0) \\
			+ \lambda_{seg} &(\frac{N_{input}}{N_{mask}} (1- \mathop{\mathbb{L}_{Dice}}(y, m_{tgt}))) \\
			\text{where}~\mathop{\mathbb{L}_{Dice}}&= 2 \frac{\sum_i y^i m_{tgt}^i}{\sum_i y^i \sum_i m_{tgt}^i}
		\end{split}
	\end{equation}
	where, $x_{src}$ and $x_{tgt}$ are the input images of the source and target domain, respectively,$\mathop{\mathbb{L}_{CE}}$ denotes the standard cross-entropy loss function, $D$ and $E$ denote the domain discriminator and the feature extractor, respectively, $y$ and $m_{tgt}$ are a segmentation result through the attention U-Net with the bidirectional C-LSTMs, and a segmentation mask of the target domain, respectively, $\theta_D$ and $\theta_{Seg}$ denote the parameters of the domain discriminator and the U-Net, respectively, including the feature extractor, attention gate, and reconstructor, $N_{input}$ and $N_{mask}$ denote the total number of pixels of the input image and mask image, respectively, and $\lambda_{seg}$ controls the relative importance of different loss terms. The higher the segmentation performance, the higher the $\mathop{\mathbb{L}_{Dice}}$; hence, we added $1- \mathop{\mathbb{L}_{Dice}}(y, m_{tgt})$ to minimize the loss function and multiplied $\frac{N_{input}}{N_{mask}}$ to weigh the small foreground. 
	
	We formulated a loss function for the discriminator to distinguish between the features of the source and target domains, as well as a loss function for the feature extractor to trick the discriminator with the target domain data, for the adversarial learning between the domain discriminator and the feature extractor. In summary, it updateed the domain discriminator and SEQ-UNET
	by minimizing (\ref{eq:domainadaptation_discriminator}) and (\ref{eq:domainadaptation_segmentation}). Both, the segmentation network and domain discriminator were optimized by training the proposed model. The feature extractor captured the domain-invariant features, to trick the domain discriminator. This helped sustain the knowledge learned from the source dataset, and improved the generalizability of the proposed model. Simultaneously, the segmentation network, including the feature extractor, the attention gates, and the reconstructor, were improved to distinguish between the background and the target object, in the field of the orbital CT images.

	\section{Experimental Results}
	\label{sec:experimentalresults}
	\subsection{Data Description and Pre-processing}
	
	\begin{table*} [t!]
			\centering
			\caption{Comparison of segmentation results on the two ORBIT CT datasets.The numbers are represented as \textit{mean$\pm$std}. }
			\label{tab:comparison}
			\begin{tabular}{c|c|c|c|c}
				\hline
				\multirow{2}{*}{Methods} & \multicolumn{2}{|c|}{PDDCA} & \multicolumn{2}{|c}{CMC-ORBIT} \\ \cline{2-5}
										 & DICE (\%) & VS (\%) & DICE (\%) & VS (\%) \\
				\hline
				H-DenseUNet \cite{li2018h}		& 64.37 $\pm$ 3.35 & 80.76 $\pm$ 4.58 & 44.28 $\pm$ 6.40 & 54.50 $\pm$ 8.40 \\
				Sensor3D \cite{novikov2018deep}	& 64.41 $\pm$ 3.12 & 79.14 $\pm$ 4.34 & 48.72 $\pm$ 8.96 & 56.76 $\pm$ 8.31 \\ 
				SEQ-UNET + DA(Ours) & \textbf{67.73 $\pm$ 3.44} & \textbf{81.69 $\pm$ 2.99} & \textbf{65.95 $\pm$ 16.23} & \textbf{68.24 $\pm$ 10.38}\\
				\hline
			\end{tabular}
		\end{table*}
		
		\begin{table*}[t!]
			\centering
			\caption{Ablation study of the proposed method on the two ORBIT CT datasets. The numbers are represented as \textit{mean$\pm$std}. }
			\label{tab:ablationstudy}
			\begin{tabular}{c|c|c|c|c}
				\hline
				\multirow{2}{*}{Methods} & \multicolumn{2}{|c|}{PDDCA} & \multicolumn{2}{|c}{CMC-ORBIT} \\ \cline{2-5}
				& DICE (\%) & VS (\%) & DICE (\%) & VS (\%) \\
				\hline
				Sensor3D \cite{novikov2018deep} & 64.41 $\pm$ 3.12 & 79.14 $\pm$ 4.34 & 48.72 $\pm$ 8.96 & 56.76 $\pm$ 8.31 \\ 
				SEQ-UNET			& 66.55 $\pm$ 2.76 & 81.12 $\pm$ 2.26 & 52.84 $\pm$ 2.55 & 63.17 $\pm$ 7.88 \\
				SEQ-UNET + TR  				& 67.26 $\pm$ 3.30 & \textbf{81.84 $\pm$ 3.83} & 42.89 $\pm$ 18.04 & 52.02 $\pm$ 7.28 \\ 
				SEQ-UNET + DA (Proposed)						& \textbf{67.73 $\pm$ 3.44} & 81.69 $\pm$ 2.99 & \textbf{65.95 $\pm$ 16.23} & \textbf{68.24 $\pm$ 10.38}\\
				\hline
			\end{tabular}
		\end{table*}
		
	For the source data, we used the Lung Image Detection Consortium (\textit{LIDC-IDRI}) dataset \cite{armato2011lung} which contained 1,018 chest CT scans and their nodule masks. We interpolated the scans to create a $1\times 1 \times 1$ mm voxel resolution, and applied histogram equalization to the entire CT scans to ensure clear visibility of the nodules. Then, we cropped the nodule from each CT scan in the size $64 \times 64 \times d$ mm, where d is the size of the nodule, and obtained the volume of interests (VOIs) of 2,536 nodules. We normalized the intensity of the slices between 0 and 1 and created the sequence data by grouping 3 slices each, to use the nodule VOIs as input for the proposed model.

	Two types of data were used as target data: optic nerve CT scans and orbital tumor CT scans. First, optic nerve CT scans were obtained from the Public Domain Database for Computational Anatomy (\textit{PDDCA}) dataset \cite{raudaschl2017evaluation}. The PDDCA dataset was used for the segmentation of several anatomical structures, such as in head and neck CT scans. The PDDCA dataset comprised of 48 CT scans, with thickness less than 3.0 mm, and an in-plane voxel resolution between 0.76 mm and 1.27 mm. The dataset also contained the manual segmentation mask for parotid glands, brainstem, optic chiasm, mandible, submandibular glands, and optic nerves. We used optic nerve masks, as we were targeting the optic area. It was similar to dealing with the LIDC-IDRI dataset, we resized the slices to make the in-plane pixel size $64 \times 64$, after interpolation and histogram equalization. Further, intensity normalization was conducted, where we binded three consecutive slices to create the input for the model.

	The second target dataset was an orbital CT scan collected from the College of Medicine, The Catholic University of Korea. We denoted the dataset by \textit{CMC-ORBIT}. This dataset contained 19 orbital CT scans that had two types of orbital tumors: Dermoid cyst and Hemangioma. The tumor masks were manually annotated by four ophthalmologists. The interpolation process was skipped, as the CT scans that were taken in high resolution, mainly had 1 or less than 0.5 mm in-plane voxel resolution and slice thickness. Instead, we implemented the window clipping method using a 48 HU clipping level and a 400 HU window. 
	To obtain orbital VOIs, we cropped the CT scans at the point where orbital tumors were contained. Additionally, we used k-mean clustering for the initial extraction of head VOIs, that were of size $h \times w \times d$ where, h and w are the height and width of the head VOIs, respectively; and d is the total number of the slices in CT scans.  
	The orbital VOIs on both sides whose size $(s, s, d)$ are obtained from head VOIs, where $s = \max(h/2, w/2)$. Once orbital tumor containing VOIs were filtered, intensity normalization was applied, and three serial slices of the orbit VOIs, with the tumor were computed for model inputs.

	\subsection{Implementation Details}
	We implemented the proposed model using Pytorch in Python. The initial weights of the network were set up through a Xavier procedure, and Adam optimizer with momentum parameters ${\beta}_{1}$=0.9 and ${\beta}_{2}$=0.9999, was employed to train both the generator and the discriminator. We trained the network for 500 epochs. Empirically determined learning rates were 0.00005 for the generator, and 0.0001 for the discriminator
	
    To apply the domain adaptation technique, we trained the attention U-Net with bidirectional C-LSTMs, using the source data in advance. Further, we initialized the attention U-Net in the source and target domain using the weights of the pre-trained model, but fixed only the weights for the attention U-Net of the source domain, to make the network extract the same feature map in the target domain.

	\subsection{Evaluation Metrics}
		
		For quantitative evaluation and comparison with the state-of-the-art algorithms, we adopted two evaluation metrics, which are used in the CT image segmentation task: Dice score (DICE) and volume similarity (VS) metrics. 
		
		\begin{equation}
			\label{eq:dice_score}
			DICE~(\%) =\frac{2\ TP}{2\ TP+FP+FN}
		\end{equation}
		\begin{equation}
			\label{eq:volume_similarity}
			VS~(\%) = 1-\frac{\left| FN-FP \right|}{2\ TP+FP+FN}
		\end{equation}
		where  $TP$, $FP$, and $FN$ denote the true positive, false positive, and false negative values, respectively, when the segmentation result $y$ and the segmentation mask $m$ are given.
		
	\begin{figure*}[!t]
    	    \centering
    	    \subfloat[]{\includegraphics[width=0.12\linewidth]{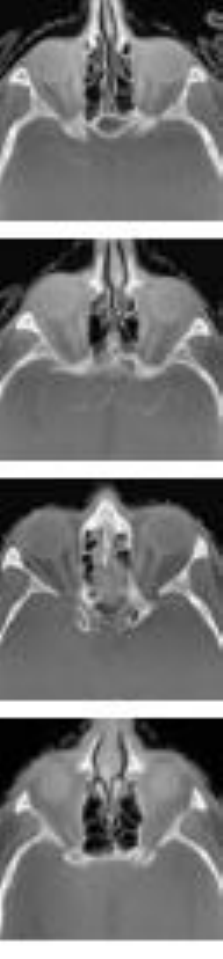}}
            \subfloat[]{\includegraphics[width=0.12\linewidth]{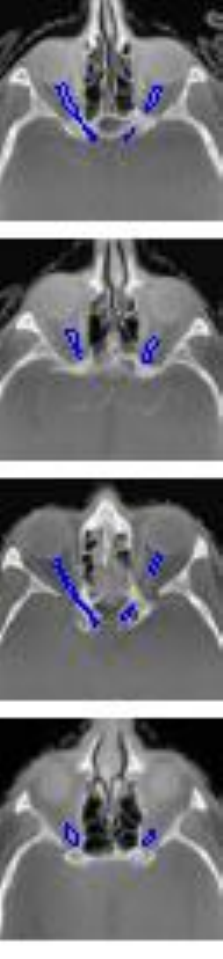}}
            \subfloat[]{\includegraphics[width=0.12\linewidth]{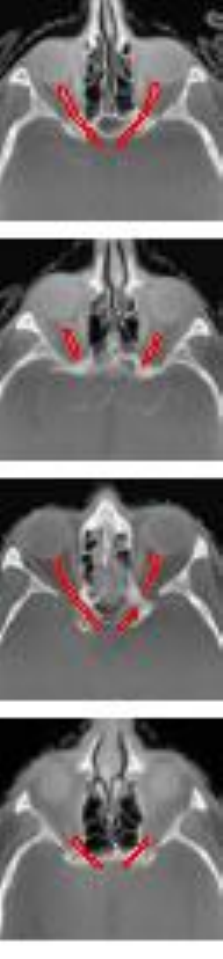}}
            \subfloat[]{\includegraphics[width=0.12\linewidth]{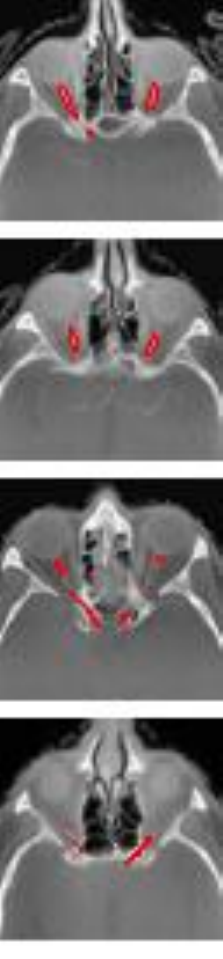}}
            \subfloat[]{\includegraphics[width=0.12\linewidth]{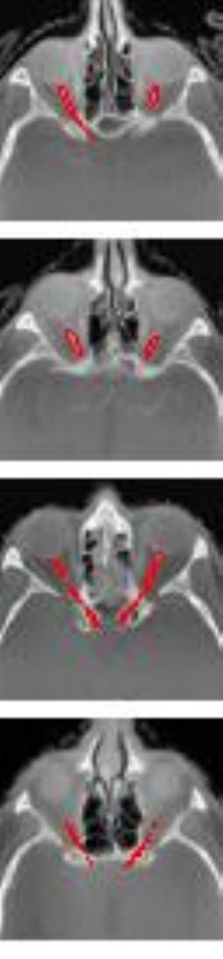}}
            \subfloat[]{\includegraphics[width=0.12\linewidth]{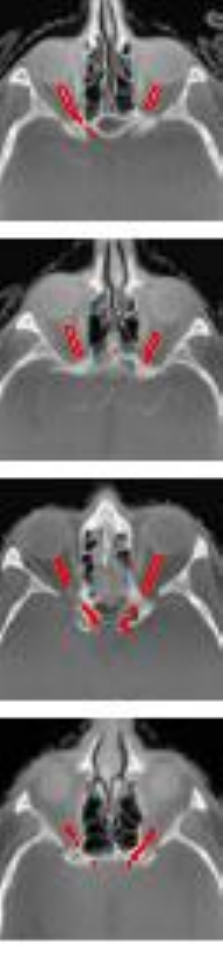}}
            \subfloat[]{\includegraphics[width=0.12\linewidth]{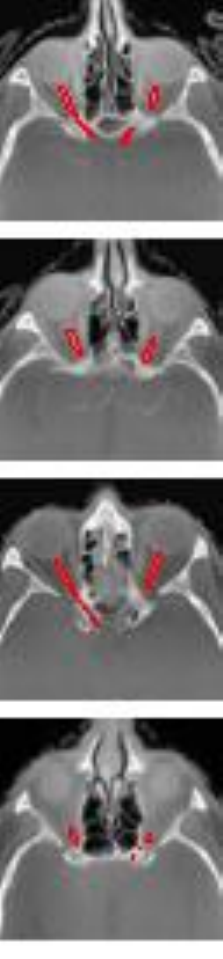}}
             
        	\caption{The visualized examples of the optic nerve segmentation on the PDDCA dataset: (a) original CT, (b) ground truth mask, predicted mask by (c) H-DenseUNet \cite{li2018h}, (d) Sensor3D \cite{novikov2018deep}, (e) SEQ-UNET, (f) SEQ-UNET + TR, (g) SEQ-UNET + DA (Ours)}
        	\label{fig:nerve_results}
	\end{figure*}
    	
	\begin{figure*}[!t]
	    \centering
	    \subfloat[]{\includegraphics[width=0.12\linewidth]{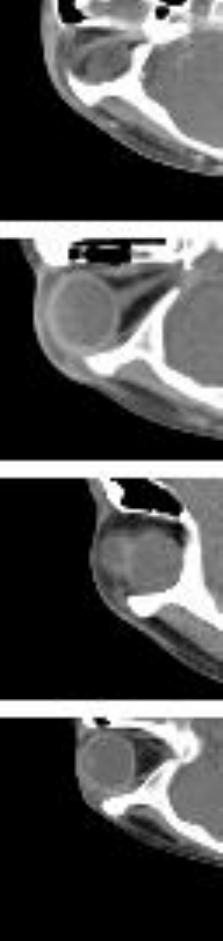}}
        \subfloat[]{\includegraphics[width=0.12\linewidth]{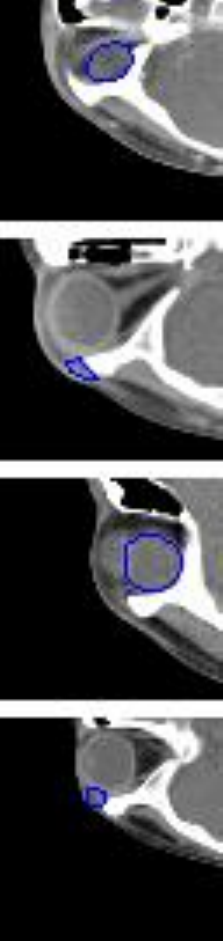}}
        \subfloat[]{\includegraphics[width=0.12\linewidth]{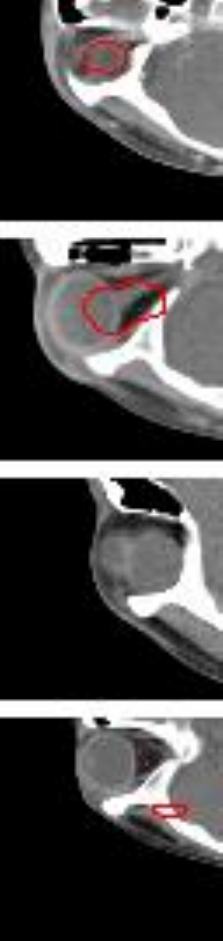}}
        \subfloat[]{\includegraphics[width=0.12\linewidth]{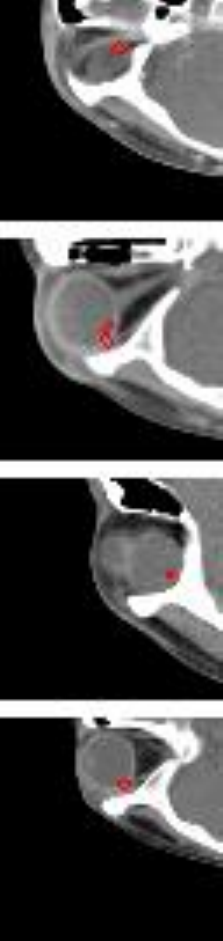}}
        \subfloat[]{\includegraphics[width=0.12\linewidth]{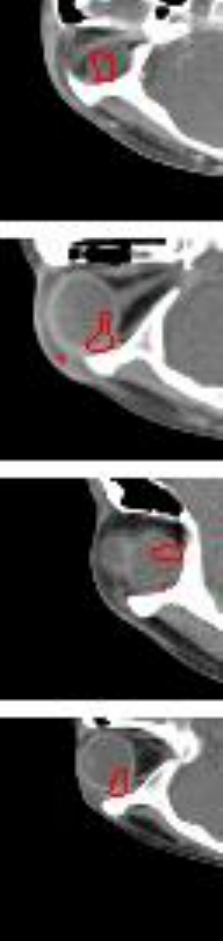}}
        \subfloat[]{\includegraphics[width=0.12\linewidth]{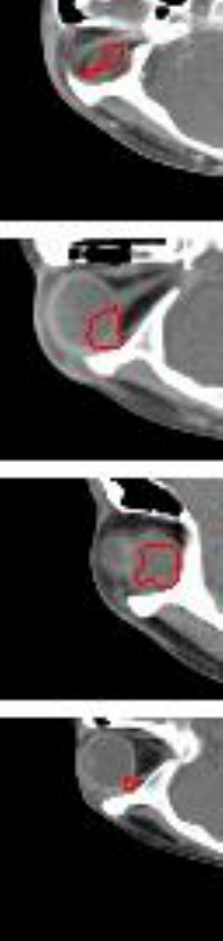}}
        \subfloat[]{\includegraphics[width=0.12\linewidth]{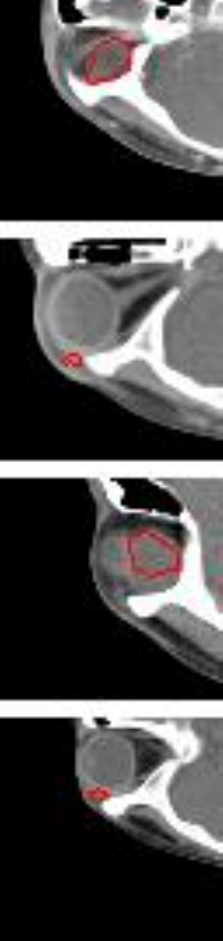}}
         
    	\caption{The visualized examples of the orbital tumor segmentation on the CMC-ORBIT dataset: (a) original CT, (b) ground truth mask, predicted masks by (c) H-DenseUNet \cite{li2018h}, (d) Sensor3D \cite{novikov2018deep}, (e) SEQ-UNET, (f) SEQ-UNET + TR, (g) SEQ-UNET + DA (Ours)} 
    	\label{fig:tumor_results}
	\end{figure*}
    	
	\subsection{Results of segmentation from small sampled orbital CTs}
		The proposed method was evaluated using the PDDCA dataset (optic nerve) and the CMC-ORBIT (orbital tumor) dataset. We emphasized that the size of the dataset must be 48 and 19, respectively, which is too small for training a deep learning model. The proposed method was evaluated and compared with the U-Net-based segmentation models: one is a H-Dense U-Net, a hybrid approach combining 2D-U-Net with 3D-U-Net architecture as proposed by Li et al. \cite{li2018h}, and the other is the Sensor3D, which integrated a 2D U-Net and bidirectional C-LSTM networks, proposed by Novikov et al. \cite{novikov2018deep}. We can show the effectiveness of the proposed attention U-Net architecture with bidirectional C-LSTMs with domain adaptation.
		
		To evaluate the proposed method, we conducted a 10-fold cross-validation procedure for the PDDCA dataset, and a 4-fold cross-validation procedure for the orbital tumor dataset. Table \ref{tab:comparison} shows the quantitative segmentation comparison results of the proposed method with the state-of-the-art methods on the two test datasets. The numbers are the mean and standard deviation values computed through cross-validation. The results show that the proposed method achieved the best performance in terms of two measurements. The score differences on the orbital tumor are significantly higher than the ones on the PDDCA. These results imply that the proposed method provides better segmentation performance, and has particularly strong advantages from a small number of training data. A few segmentation results of PDDCA and CMC-ORBIT are shown in Fig. \ref{fig:nerve_results} and Fig. \ref{fig:tumor_results}.

	\begin{figure}[!t]
    		\centering
    	    \subfloat[Dermoid cyst blurred in black]{\includegraphics[width=0.75\linewidth]{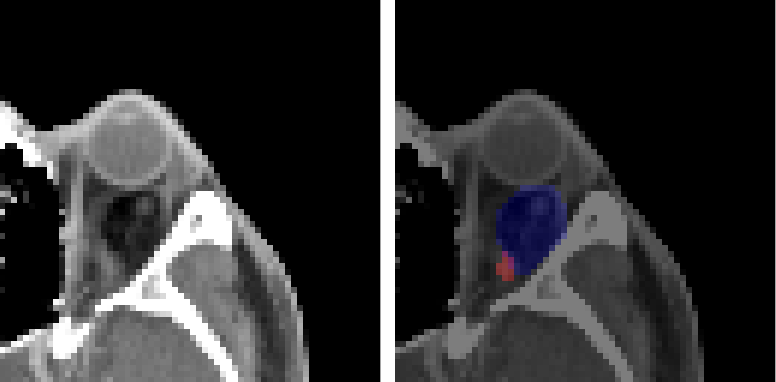}} \\
            \subfloat[Dermoid cyst close to  tissues]{\includegraphics[width=0.75\linewidth]{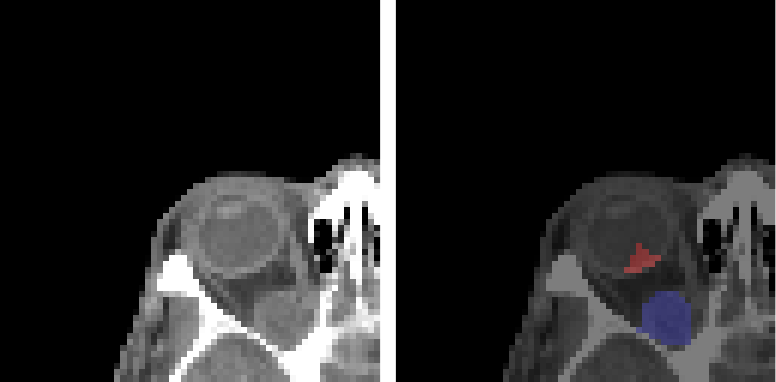}} \\
            \subfloat[Dermoid cyst outside of orbit]{\includegraphics[width=0.75\linewidth]{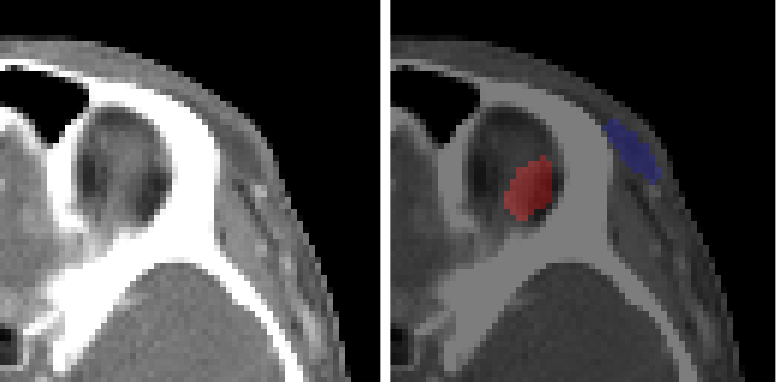}} \\
            \subfloat[Hemangioma]{\includegraphics[width=0.75\linewidth]{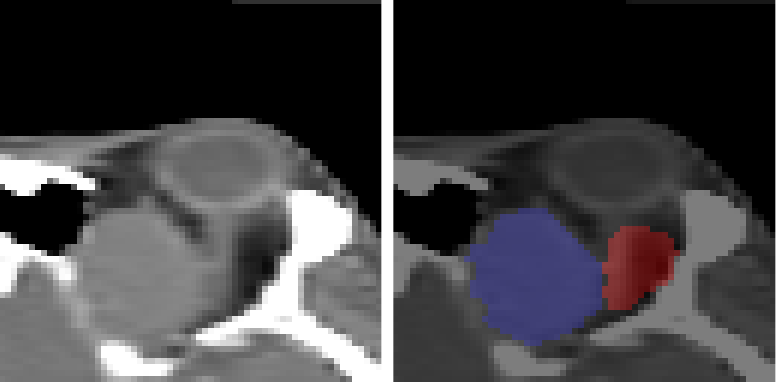}} \\
            \subfloat[Pleomorphic Adenoma]{\includegraphics[width=0.75\linewidth]{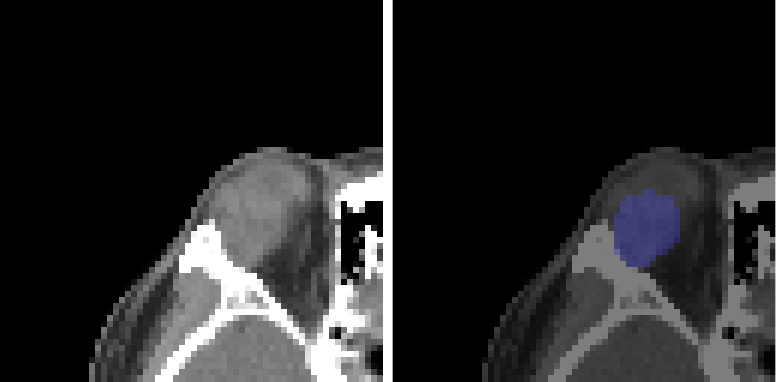}}
    		\caption{Failed cases in testing additional orbital tumor CTs (ground truth masked in blue and predicted segmentation in red)}
    		\label{fig:limits}
    	\end{figure}

	\subsection{Model Ablation Study}
		To evaluate the effectiveness of our proposed attention U-Net architecture with bidirectional C-LSTMs and domain adaptation, we conducted the ablation study. 
		We established the performance of the Sensor3D as a baseline, because Sensor3D is an architecture with bidirectional C-LSTMs. \textit{SEQ-UNET} was the proposed U-Net architecture only with the target task dataset without domain adaptation. For the comparison between domain adaptation and transfer learning, \textit{SEQ-UNET + TR} is a transfer learning method by fixing weights of the part of the network and fine-tuning the rest of them. 
		Table \ref{tab:ablationstudy} shows the segmentation results by ablation study of the proposed method. Firstly, the \textit{SEQ-UNET + TR} provided better performance than the baseline, in terms of the two metrics on the datasets. It indicates that the proposed attention of the U-Net with bidirectional C-LSTMs, inherits the advantage of the Sensor3D, and improved the performance of segmentation, by adopting an approach where, the attention gates before the skip connections in the U-Net architecture. 
		
		Further, as compared to \textit{Transfer learning}, the proposed domain adaptation outperformed in both datasets. It also showed that transfer learning was not as effective in the CMC-ORBIT dataset, when the dataset was extremely small, a cross-domain dataset is given. The effectiveness of domain adaptation with the domain discriminator is shown in this comparison.   
		
		 The effects of domain adaption have been visualized in the Figs. \ref{fig:nerve_results} and \ref{fig:tumor_results}. Application of domain adaptation to \textit{SEQ-UNET}, improved DICE and VS much more than transfer learning as shown in Table \ref{tab:ablationstudy}.  When sensitivity and specificity were measured in the CMC-ORBIT, we found that \textit{SEQ-UNET + DA} improved sensitivity, while \textit{SEQ-UNET + TR} exhibited a performance similar to \textit{SEQ-UNET} as shown in Table \ref{tab:sensitivity_specificity}.
	
	\subsection{Limitations}
	    After testing the given CMC-ORBIT dataset, we conducted testing on additional orbital CT data and analyzed the failed case of the proposed methods as shown in Fig.~\ref{fig:limits}. In the case of a Dermoid cyst, when the tumor was blurred in dark, it was not detected. Also, the model does not provide well-segmented results in some specific cases, where the tumor is very small, close to unmarked tissues, and located out of orbit. In the case of Hemangioma, the trained model often failed to detect and segment partially, even though Hemangioma is highly similar to Dermoid cyst in a CT scan. Subsequently, Pleomorphic Adenoma, heterogeneous to Dermoid cyst, is undetectable by the model trained with Dermoid cyst and Hemangioma.

		\begin{table} [!t]
			\caption{Performance comparison for the orbital tumor segmentation}
			\label{tab:sensitivity_specificity}
			\centering
			\begin{tabular}{c|c|c|c}
				\hline
				 & SEQ-UNET & SEQ-UNET + TR & SEQ-UNET + DA \\ 
				 \hline
				sensitivity & 0.3309 & 0.3455 & 0.5350 \\
				specificity & 0.9985 & 0.9966 & 0.9977 \\
				\hline
			\end{tabular}
		\end{table}

	\section{Conclusions}
	\label{sec:conclusions}
	
	We have proposed SEQ-UNET + DA, which is an attention U-Net architecture with bidirectional C-LSTMs, and a U-Net architecture-based domain adaptation, for segmentation of small sampled CT images. When a small number of orbital CT images, and masks of the optic nerve and orbital tumor are given, the performance of 2D and 3D U-net is poor. However, SEQ-UNET improved the segmentation performance with lower computation. In addition, the proposed domain adaptation further improved the segmentation of the optic nerve and orbital tumors, even though cross-domain adaptation that was applied using the lung nodule dataset. We expect that the proposed method will help, applications of deep learning-based segmentation, to new and rare diseases where only a limited dataset is available. As for future studies, we are looking into the methods of enhancing domain adaptation, such as feature map regularization, for medical image analysis.


	\bibliographystyle{IEEEtran}
	\bibliography{Supervised_Segmentation_with_Domain_Adaptation}
	
\end{document}